\documentclass{article}
\usepackage{amssymb}
\usepackage{amsmath}

\begin{document}

\title{\bf Generalized zeta functions, shape invariance and one-loop
corrections to quantum Kink masses}
\author{S. Rafiei, S. Jalalzadeh\thanks{email:
s-jalalzadeh@sbu.ac.ir} and K. Ghafoori
Tabrizi\thanks{email: k-tabrizi@sbu.ac.ir}
\\ {\small Department of Physics, Shahid Beheshti University, Evin, Tehran
19839, Iran}}
\maketitle
\begin{abstract}
\vspace{0.5cm} We present a method to calculate the One-loop mass
correction to  Kinks mass in a $(1+1)$-dimensional field
theoretical model in which the fluctuation potential
$V^{\prime\prime}(\phi_c)$ has shape invariance property. We
use the  generalized zeta function regularization method to
implement our setup for describing quantum kink states.
\end{abstract}
\vspace{1cm}
\section{Introduction}

The quantum corrections to  the mass of classical topological defects plays
an important role in the semi-classical approach to the quantum
field theory [1]. Computation of quantum energies around
classical configurations in $(1+1)$-dimensional kinks has been
developed in [2] by using topological boundary conditions,
derivative expansion method [3], scattering phase shift technique
[4], mode regularization approach [5], zeta-function
regularization technique [6] and also dimensional regularization method [7].
In this paper we will give a derivation of the one-loop
renormalized  kink quantum mass correction in a
$(1+1)$-dimensional scalar field theory model using
generalized zeta function method for those potentials where the
 fluctuation potential, $V''(\phi_{c})$ has the shape
invariance property. This kind of potential is most occurrent in
different fields of physics, particularly in quantum gravity and
cosmology. What makes these potentials so important is that they
posses a shape invariant operator in their prefactor, making
corrections of these kind of potentials exact by the
heat kernel method.

Consider the (1+1)-dimensional scalar field theory, where
classical dynamics is described by following action functional for
scalar $\phi(x)$ with potential $V(\phi)$
\begin{eqnarray}\label{1}
S[\phi]=\int d^2x\left[\frac{1}{2}\phi_{,\mu} \phi^{,\mu}-V(\phi)\right],
\end{eqnarray}
where a semicolon denotes ordinary derivation in two dimensional
Minkowskian space-time. We shall consider the following examples
\begin{eqnarray}\label{2}
V(\phi) = \left\{\begin{array}{ll} \frac{m^4}{\lambda}\left(1 -
\cos(\frac{\sqrt{\lambda}\phi}{m})\right);  \,\,\,Sine-Gordon (SG)\,\,
 model \vspace{.5 cm} \\ \frac{m^4}{4\lambda}\left((\frac{\sqrt{\lambda}}{m}\phi)^2 -
1\right)^2;\,\,\,\,\,\phi^4-model.
\end{array}\right.
\end{eqnarray}
These two kinds of potentials are tex-book cases. The method that
we use can easily be developed to include all potentials that have shape
invariant property. There are a number of papers where this kind
of potentials are discussed, see for example [8]. We know that
mass $m$ and coupling constant $\lambda$ may be scaled out so
that  semi-classical expansion is an expansion in powers of
$\frac{\hbar \lambda}{m^2}$ and therefore we can set $m = \lambda
= 1$. Classical static kink-antikink solutions $\phi_{c}(x)$
satisfy
\begin{eqnarray}\label{3}
\phi_{c}' = \pm\sqrt{2V(\phi_{c})}.
\end{eqnarray}
Therefore for potentials in (2), the classical solutions are
\begin{eqnarray}\label{4}
\phi_c =  \left\{ \begin{array}{ll}4\arctan(e^{x}); \hspace{.5 cm}
SG\,\,model \\\\
\tanh(\sqrt{2}x); \hspace{.5 cm} \phi^4-model.
\end{array}\right.
\end{eqnarray}
As we will see in the next section, to compute  mass
corrections we need the fluctuation potential $V''(\phi_c(x))$.
For  cases listed above, we have
\begin{eqnarray}\label{5}
V''(\phi_c) = \left\{ \begin{array}{ll}1 - \frac{2}{\cosh^2x};
\hspace{.5 cm} SG\,\, model \\\\
4 - \frac{6}{\cosh^2x}; \hspace{.5 cm} \phi^4-model.
\end{array}\right.
\end{eqnarray}
In the next section we quote briefly the generalized zeta function
method. We then use  shape invariance property of
potentials (5) and the heat kernel method to obtain  quantum
corrections to  kink masses.
\section{Semi-Classical Quantum Kink States}
Classical configuration space is found by
static configuration $\Phi(x)$, so that the energy functional
corresponding to  classical action functional (1)
\begin{eqnarray}\label{6}
E[\Phi]=\int dx\left[\frac{1}{2}\Phi_{,\mu}
\Phi^{,\mu}+V(\Phi)\right],
\end{eqnarray}
is finite. One can describe quantum evolution in  Schrodinger
picture by the following functional equation
\begin{eqnarray}\label{7}
i\hbar\frac{\partial}{\partial t}\Phi[\phi(x),t]=H\Phi[\phi(x),t],
\end{eqnarray}
so that  quantum Hamiltonian operator is given by
\begin{eqnarray}\label{8}
H=\int
dx\left[-\frac{\hbar^2}{2}\frac{\delta}{\delta\phi(x)}\frac{\delta}{\delta\phi(x
)}+E[\phi]\right].
\end{eqnarray}
In  field representation  matrix elements of evolution
operator are given by
\begin{eqnarray}\label{9}
G(\phi^{(f)}(x),\phi^{(i)}(x),T)=\langle\phi^{(f)}|e^{-\frac{iT}{\hbar}H}|\phi^{
(i)}\rangle
=\int D[\phi(x,t)]\exp{(\frac{-i}{\hbar}S[\phi])},
\end{eqnarray}
where the initial conditions are those of static solutions kink  of  classical equations where $\phi^{(i)}(x,0)=\phi_{k}(x)$, $\phi^{(f)}(x,T)=\phi_{k}(x)$. In  semi-classical picture, we
are interested in loop expansion for evolution operator up to the  first quantum correction
\begin{eqnarray}\label{10}
G(\phi^{(f)}(x),\phi^{(i)}(x),\beta)=\exp{(-\frac{\beta}{\hbar}E[\phi_k])}
Det^{-\frac{1}{2}}\left[-\partial^2_{\tau}+P\Delta\right](1+{\cal
O}(\hbar)),
\end{eqnarray}
where we use analytic continuation to Euclidean time,
$t=-i\tau$,$T=-i\beta$, and $\Delta$ is the differential
operator
\begin{eqnarray}\label{11}
 \Delta=-\frac{d^2}{d
 x^2}+\frac{d^2V}{d\phi^2}\mid_{\phi=\phi_k},
\end{eqnarray}
 P is the projector over
the strictly positive part of spectrum of $\Delta$
\begin{eqnarray}\label{12}
\Delta\xi_n(x)=\omega_n^2\xi_n(x),\,\,\,\,\omega^{2}_{n}\,\,
\epsilon \,\, Spec(\Delta)= Spec(P\Delta)+\{0\}.
\end{eqnarray}
We write  functional determinant in  the form
\begin{eqnarray}\label{13}
Det\left[-\frac{\partial^2}{\partial\tau^2}+\Delta\right]=\prod_{n}det\left[-\frac{\partial^2}{
\partial\tau^2}+\omega^{2}_{n}\right].
\end{eqnarray}
All  determinants in  infinite product correspond to harmonic
oscillators of frequency $\omega_n$. On the other hand, it is well known
that \cite{Feyn}
\begin{eqnarray}\label{13a}
\begin{array}{cc}
det\left(-\frac{\partial^{2}}{\partial \tau^2}+\omega^2_n \right)^{-\frac{1}{2}}
= \prod_{j=1}^N \left(\frac{j^2\pi^2}{\beta^2}+\omega_n^2\right)^{-\frac{1}{2}}\\
\\
= \prod_j\left(\frac{j^2\pi^2}{\beta^2}\right)^{-\frac{1}{2}}\prod_j\left(1+\frac{\omega^2_n\beta^2}{j^2\pi^2}\right)^{-\frac{1}{2}}.
\end{array}
\end{eqnarray}
The first product dose not depend on $\omega_n$ and combines with the Jacobian
and other factors we have collected into a single constant. The second factor
has the limit $\left[\frac{\sinh(\omega_n\beta)}{\omega_n\beta}\right]^{-\frac{1}{2}}$,
and thus, with an appropriate normalization, we obtain for large $\beta$
\begin{eqnarray}\label{14}
G(\phi^{(f)}(x),\phi^{(i)}(x),\beta)\cong
\exp{(-\frac{\beta}{\hbar}E[\phi_k])}\prod_{n}(\frac{\omega_n}{\pi\hbar})^{\frac{1
}{2}}\exp{\left(-\frac{\beta}{2}\sum_{n}\omega_n(1+{\cal
O}(\hbar))\right)}
\end{eqnarray}
where eigenvalues in the kernel of $\Delta$ have been
excluded. Interesting eigenenergy wave functionals
\begin{eqnarray}\label{15}
H\Phi_j[\phi_k(x)]=\varepsilon_j\Phi_j[\phi_{k}(x)]
\end{eqnarray}
we have an alternative expression for $G_E$ for
$\beta\rightarrow\infty$.
\begin{eqnarray}\label{16}
G(\phi^{(f)}(x),\phi^{(i)}(x),\beta)\cong
\Phi^{*}_{0}[\phi_k(x)]\Phi_{0}[\phi_k(x)]\exp{(-\beta\frac{\varepsilon_0}{\hbar
})},
\end{eqnarray}
and, therefore, from (14) and (16) we obtain
\begin{eqnarray}\label{17}
\varepsilon_0=E[\phi_k]+\frac{\hbar}{2}\sum_{\omega^{2}_{n}>0}\omega_n+{\cal
O}(\hbar),
\end{eqnarray}
\begin{eqnarray}\label{18}
|\Phi_0[\phi_k(x)]|^2=Det^{\frac{1}{4}}\left[\frac{P\Delta}{\pi^2\hbar^2}\right],
\end{eqnarray}
as the Kink ground state energy and wave functional up to One-Loop
order.\\
 If we define
the generalized zeta function
\begin{eqnarray}\label{19}
\zeta_{P\triangle}=Tr(P\Delta)^{-s}=\sum_{\omega^2_n>0}\frac{1}{(\omega^2_n)^
s},
\end{eqnarray}
 associated to  differential operator $P\triangle$, then
\begin{eqnarray}\label{20}
\varepsilon_0^k=E[\phi_k]+\frac{\hbar}{2}Tr(P\Delta)^{\frac{1}{2}}+{\cal
O }(\hbar^2)
=E[\phi_k]+\frac{\hbar}{2}\zeta_{P\Delta}(-\frac{1}{2})+{\cal
O} (\hbar^2).
\end{eqnarray}
The eigenfunction of $\Delta$ is a basis for  quantum
fluctuations around kink background, therefore sum of the
associated zero-point energies encoded in
$\zeta_{P\Delta}(-\frac{1}{2})$ in (20) is infinite. According
to zeta function regularization procedure, energy and mass
renormalization prescription,   renormalized kink energy in
semi-classical limit becomes [9]
\begin{eqnarray}\label{21}
\varepsilon^k(s)=E[\phi_k]+\Delta M_k +{\cal O}(\hbar^2)
=E[\phi_k]+\lim_{s\rightarrow\frac{-1}{2}}[\delta_1\varepsilon^k(s)+\delta_
2^k\varepsilon(s)]+{\cal
O}(\hbar^2),
\end{eqnarray}
where
\begin{eqnarray}\label{22}
\left\{
\begin{array}{rr} \delta_1\varepsilon^k(s)=
\frac{\hbar}{2}\mu^{2s+1}[\zeta_{P\Delta}(s)-\zeta_{\nu}(s)],\hspace{4.3cm}
\\\\ \delta_2\varepsilon^k(s)=
\lim_{L\rightarrow\infty}\frac{\hbar}{2L}\mu^{2s+1}\frac{\Gamma(s+1)}{\Gamma(s)
}
\zeta_\nu(s+1)\int_{-\frac{L}{2}}^{\frac{L}{2}}
dx\left[\frac{d^2V}{d\phi^2}|_{\phi_k}-\frac{d^2V}{d\phi^2}|_{\phi\nu}\right].
\end{array}
\right.
\end{eqnarray}
Here $\phi_\nu$ is a constant minimum of potential $V(\phi)$, $E$ is corresponding classical energy where $\mu$ has the unit  $length^{-1}$ dimension, introduced to make the terms in (\ref{22}) homogeneous from
a dimensional point of view and $\zeta_\nu$ denoted zeta
function associated with vacuum $\phi_v$.\\
Now we explain very briefly how one can calculate  zeta
function of an operator though  heat kernel method. We
introduce  generalized Riemann zeta function of operator A
by
\begin{eqnarray}\label{23}
\zeta_{A}(s)=\sum_{n}\frac{1}{|\lambda_n|^s},
\end{eqnarray} where
$\lambda_n$ are eigenvalues of operator $A$.
  On the other hand, $\zeta_A(s)$ is
 the Mellin transformation of heat kernel $G(x,y,t)$ which satisfies
 the following heat diffusion equation
 \begin{eqnarray}\label{24}
 A G(x,y,t)=-\frac{\partial}{\partial t}G(x,y,t),
 \end{eqnarray}
 with an initial condition $G(x,y,0)=\delta(x-y)$. Note that
 $G(x,y,t)$ can be written in terms of its spectrum
 \begin{eqnarray}\label{25}
 G(x,y,t)=\sum_{n}e^{-\lambda_n t}\psi_n^{*}(x)\psi_n(y),
 \end{eqnarray}
 and as usual, if the spectrum is continues, one should integrate it. From relation (17), it is clear that
 \begin{eqnarray}\label{26}
 \zeta_A(s)=\frac{1}{\Gamma(s)}\int_{0}^{\infty}d\tau\tau^{s-1}\int_{-\infty}^{
 \infty}G(x,x,\tau)dx.
 \end{eqnarray}
 Hence, if we know the associated Green function of an operator,
  we can calculate  generalized zeta function corresponding
 to that operator. In  the next sections we calculate  the Green
 function of $\phi^4$-model and SG via shape invariance
 property and there, by using equations (\ref{24}), (\ref{25}) and (\ref{26})
  we will  obtain  one-loop corrections to quantum  kink masses.

\section{Quantum Mass of SG and $\phi^4$-models}
In this section we calculate one-loop quantum mass of these two potentials.
According to the previous section the second derivative of these potentials
at the Kink solution can be written as
\begin{eqnarray}\label{27}
U(x)=l^2-\frac{l(l+1)}{cosh^2(x)},
\end{eqnarray}
so that for $l=1$ and $l=2$ we obtain SG and $\phi^4$-model second derivative
potentials respectively. Therefor the operator (\ref{11}) which acts on the
eigenfunctions becomes
\begin{eqnarray}\label{28}
\Delta_l = -\frac{d^2}{dx^2} +l^2-\frac{l(l+1)}{cosh^2(x)}.
\end{eqnarray}
Also the operator acting on the vacuum has the following form
\begin{eqnarray}\label{29}
\Delta_l(0)=-\frac{d^2}{dx^2}+l^2.
\end{eqnarray}
In the reminding of this section, to obtaining the spectrum of (\ref{28}) we will use the shape invariance property. First we review briefly concepts
that we will use.\\
Consider the following one-dimensional bound-state Hamiltonian
\begin{eqnarray}\label{30}
H= -\frac{d^2}{dx^2}+U(x), \hspace{2cm} x\in I \subset \mathbb{R}
\end{eqnarray}
where $I$ is the domain of  $x$ and $U(x)$ is a real function of $x$,
which can be singular only in the boundary points of the domain. Let us denote
by $E_n$ and $\psi_n(x)$ the eigenvalues and eigenfunctions of $H$ respectively.
We use factorization method which consists of writing Hamiltonian as the product
of two first order mutually adjoint differential operators $A$ and $A^\dagger$.
If the ground state eigenvalue and eigenfunctions are known, then one can
factorize Hamiltonian (\ref{30}) as
\begin{eqnarray}\label{31}
H=A^\dagger A +E_0,
\end{eqnarray}
where $E_0$ denotes the ground-state eigenvalue,
\begin{eqnarray}\label{32}
\begin{array}{lll}
A=\frac{d}{dx}+W(x),\\
\\
A^\dagger=-\frac{d}{dx}+W(x),
\end{array}\
\end{eqnarray}
and
\begin{eqnarray}\label{33}
W(x)=-\frac{d}{dx}\ln(\psi_0).
\end{eqnarray}
Supersymmetric quantum mechanics (SUSY QM) begins with a set of two matrix
operators, known as supercharges
\begin{eqnarray}\label{34}
Q^+ = \begin{pmatrix}0& A^\dagger \\
0 & 0 \\
\end{pmatrix}, \hspace{2cm} Q^-=\begin{pmatrix}0 & 0 \\
A & 0\\
\end{pmatrix}.
\end{eqnarray}
This operators form the following superalgebra \cite{Cooper}
\begin{eqnarray}\label{35}
\{Q^+,Q^-\}=H_{SS}, \hspace{1cm} [H_{SS},Q^{\pm}]=(Q^\pm)^2=0,
\end{eqnarray}
where SUSY Hamiltonian $H_{SS}$ is defined as
\begin{eqnarray}\label{36}
H_{SS}=\begin{pmatrix}A^\dagger A & 0 \\
0 & AA^\dagger \\
\end{pmatrix} =\begin{pmatrix}H_1 & 0 \\
0 & H_2 \\
\end{pmatrix}.
\end{eqnarray}
In terms of the Hamiltonian supercharges
\begin{eqnarray}\label{37}
\begin{array}{ccc}
Q_1=\frac{1}{\sqrt{2}}(Q^++Q^-),\\
\\
Q_2=\frac{1}{\sqrt{2i}}(Q^+-Q^-),
\end{array}
\end{eqnarray}
the superalgebra takes the form
\begin{eqnarray}\label{38}
\{Q_i,Q_j\}=H_{SS}\delta_{ij}, \hspace{0.5cm}[H_{SS},Q_i]=0, \hspace{0.5cm}
i,j=1,2.
\end{eqnarray}
The operators $H_1$ and $H_2$
\begin{eqnarray}\label{39}
\begin{array}{cc}
H_1= A^\dagger  A = -\frac{d^2}{dx^2} +U_1=-\frac{d^2}{dx^2}+W^2-\frac{dW}{dx},\\
\\
H_2=AA^\dagger = -\frac{d^2}{dx^2} + U_2 =-\frac{d^2}{dx^2}+W^2+\frac{dW}{dx},
\end{array}
\end{eqnarray}
are called SUSY partner Hamiltonians and the function $W$ is called the superpotential.
Now, let us denote by $\psi^{(1)}_{\,\,\, l}$ and $\psi^{(2)}_{\,\,\, l}$
the eigenfunctions of $H_1$ and $H_2$ with eigenvalues $E^{(1)}_{l}$ and $E^{(2)}_l$, respectively. It is easy to see that the eigenvalues of the
above Hamiltonians are positive and isospectral, i.e., they have almost the
same energy eigenvalues, except for the ground state energy of $H_1$. According
to the \cite{Cooper}, their energy spectra are related as
\begin{eqnarray}\label{40}
\begin{array}{cccc}
E_l=E^{(1)}_l+E_0, & E^{(1)}_0=0, & \psi_l=\psi^{(1)}_l,& l=0,1,2,.., \\
\\
E^{(2)}_l=E^{(1)}_{l+1},\\
\\
\psi^{(2)}_l = [E^{(1)}_{l+1}]^{-\frac{1}{2}}A\psi^{(1)}_{l+1},\\
\\
\psi^{(1)}_{l+1} = [E^{(2)}_{l}]^{-\frac{1}{2}}A^\dagger\psi^{(2)}_{l}.
\end{array}
\end{eqnarray}
Therefor if the eigenvalues and eigenfunctions of $H_1$ were known, one could
immediately derive the spectrum of $H_2$. However the above relations only
give the relationship between the eigenvalues and eigenfunctions of the two
partner Hamiltonians. A condition of an exactly solvability is known as the
shape invariance condition. This condition means the pair of SUSY partner
potentials $U_{1,2}(x)$ are similar in shape and differ only in the parameters
that appears in them \cite{Gen},
\begin{eqnarray}\label{41}
U_2(x;a_1)=U_2(x;a_2)+{\cal R}(a_1),
\end{eqnarray}
where $a_1$ is a set of parameters and $a_2$ is a function of $a_1$. Then
the eigenvalues of $H_1$ are given by
\begin{eqnarray}\label{42}
E^{(1)}_l={\cal R}(a_1)+{\cal R}(a_2)+...+{\cal R}(a_l),
\end{eqnarray}
and the corresponding eigenfunctions are
\begin{eqnarray}\label{43}
\psi_l=\prod^{l}_{m=1}\frac{A^\dagger(x;a_m)}{\sqrt{E_m}}\psi_0(x;a_{l+1}).
\end{eqnarray}
The shape invariance condition (\ref{41}) can be rewritten in terms of the
factorization operators defined in equation (\ref{32})
\begin{eqnarray}\label{44}
A(x;a_1)A^\dagger(x;a_1) = A^\dagger(x;a_2) A(x;a_2)+{\cal R}(a_1),
\end{eqnarray}
where $a_2=f(a_1)$.\\
Now we are ready to obtain spectra of $\Delta_l$ operator defined in (\ref{28}). For a given eigenspectrum of $E_l$, we introduce the following
factorization operators
\begin{eqnarray}\label{45}
\begin{array}{cc}
A_l=\frac{d}{dx}+l\tanh(x),\\
\\
A^\dagger_l=-\frac{d}{dx}+l\tanh(x),
\end{array}
\end{eqnarray}
the operator $\Delta_l$ can be factorized as
\begin{eqnarray}\label{46}
\begin{array}{cc}
 A^\dagger_l(x)A_l(x)\psi^{(1)}_n(x)=E^{(1)}_n\psi^{(1)}_n(x),\\
 \\
 A_l(x)A^\dagger_l(x)\psi^{(2)}_n(x)=E^{(2)}_n\psi^{(2)}_n(x).
 \end{array}
 \end{eqnarray}
 Therefor for a given $l$, its first bounded excited state can be obtained
 from the ground state of $l-1$ and consequently the excited state $m$ of
 a given $l$, $\psi_{l,m}(x)$, using (\ref{43}) can be written as
 \begin{eqnarray}\label{47}
 \psi_{l,m}(x) = \sqrt{\frac{2(2m-1)!}{\Pi_{j=1}^mj(2l-j)}}\frac{1}{2^m(m-1)!}A^\dagger_l(x)
 A^\dagger_{l-1}(x)...A^\dagger_{m+1}(x)\frac{1}{\cosh^m(x)},
 \end{eqnarray}
 with eigenvalue $E_{l,m}=m(2l-m)$. Obviously its ground state with $E_{l,0}=0$
 is given by $\psi_{l,0}\propto \cosh^{-l}(x)$. Also its continuous spectrum
 consists of
 \begin{eqnarray}\label{48}
 \psi_{l,k}(x)=\frac{A^\dagger_{l}(x)}{\sqrt{k^2+l^2}}\frac{A^\dagger_{l-1}(x)}{\sqrt{k^2+(l-1)^2}}
 ...\frac{A^\dagger_{1}(x)}{\sqrt{k^2+1}}\frac{e^{ikx}}{\sqrt{2\pi}},
 \end{eqnarray}
 with eigenvalues $E_{l,k}=l^2+k^2$ with following  normalization condition  \begin{eqnarray}\label{49}
 \int_{-\infty}^\infty \psi^*_{l,k}(x)\psi_{l,k'}(x)dx=\delta(k-k').
 \end{eqnarray}
 Therefor, using equations (\ref{24}), (\ref{25}), (\ref{47}) and (\ref{48})
 we find
 \begin{eqnarray}\label{50}
 G_{\Delta_l(0)}(x,y,\tau)=\frac{e^{-l^2\tau}}{2\sqrt{\pi \tau}}e^{-(x-y)^2/4\tau},
 \end{eqnarray}
 and
 \begin{eqnarray}\label{51}
 \begin{array}{cc}
 G_{\Delta_l}(x,y,\tau)=\sum_{m=1}^{l-1}\psi^*_{l,m}(x)\psi_{l,m}(y)e^{-m(2l-m)\tau}\\
 \\
 +\int_{-\infty}^\infty  \frac{dk}{2\pi}\frac{e^{-(l^2+k^2)\tau}}{\prod_{m=1}^l(k^2+m^2)}\left(\prod_{m=1}^lA^\dagger_m(x)e^{ikx}\right)^*
\left(\prod_{m=1}^lA^\dagger_m(y)e^{iky}\right).
\end{array}
\end{eqnarray}
Hence, for $l=1$ (SG), according to  (\ref{26}) it is easy to show that
\begin{eqnarray}\label{52}
\xi_{P\Delta_1}(s) - \xi_{\Delta_1(0)}(s)=-\frac{1}{\pi}\int_{-\infty}^\infty
\frac{dk}{(k^2+1)^{s+1}}=-\frac{1}{\sqrt{\pi}}\frac{\Gamma(s+\frac{1}{2})}{\Gamma(s+1)}.
\end{eqnarray}
Consequently according to the (\ref{22}) the first correction term to the
kink quantum mass of SG becomes
\begin{eqnarray}\label{53}
\delta_1\varepsilon^k(s)=\frac{\hbar}{2}\mu^{2s+1}\left[\xi_{P\Delta_1}(s)-\xi_{\Delta_1(0)}(s)\right]=-\frac{\hbar}{2\sqrt{\pi}}\mu^{2s+1}\frac{\Gamma(s+\frac{1}{2})}{\Gamma(s+1)}.
\end{eqnarray}
The second correction term  is also  given by
\begin{eqnarray}\label{54}
\begin{array}{cc}
\delta_2\varepsilon^k(s)=\lim _{L\rightarrow \infty}\frac{\hbar}{2L}\mu^{2s+1}\frac{\Gamma(s+\frac{1}{2})}{\Gamma(s)}\xi_{\Delta_1(0)}(s+1)\int_{-\frac{L}{2}}^\frac{L}{2}\left(1-\frac{2}{\cosh^2(x)}-1\right)dx\\
\\
=-\lim_{L\rightarrow\infty}\frac{\hbar}{2L}\mu^{2s+1}\frac{\Gamma(s+1)}{\Gamma(s)}\frac{L}{2\sqrt{\pi}\Gamma(s+1)}\Gamma(s+\frac{1}{2})2\tanh(\frac{L}{2})=\\
\\
-\frac{\hbar}{\sqrt{\pi}}\mu^{2s+1}\frac{\Gamma(s+\frac{1}{2})}{\Gamma(s)}
\end{array}
\end{eqnarray}
Therefore the corrected mass for SG kink is derived from
\begin{eqnarray}\label{55}
\varepsilon^k(s)=E[\phi_k]+\lim_{s\rightarrow
-\frac{1}{2}}\left[\delta_1\varepsilon^k(s)+\delta_2\varepsilon^k(s)\right],
\end{eqnarray}
Using  the variable $\alpha=s+\frac{1}{2}$, functions
$\delta_1\varepsilon^k(s)$ and $\delta_2\varepsilon^k(s)$
can be written in the following form
\begin{eqnarray}\label{56}
\left\{\begin{array}{rr}
\delta_1\varepsilon^k(\alpha)= -\frac{\hbar}{2\sqrt{\pi}}\mu^{2s}\frac{\Gamma(\alpha)}{\Gamma(\alpha+\frac{1}{2})},
\\\\\
\delta_1\varepsilon^k(\alpha)= -\frac{\hbar}{\sqrt{\pi}}\mu^{2s}\frac{\Gamma(\alpha)}{\Gamma(\alpha-\frac{1}{2})},
\end{array}
\right.
\end{eqnarray}
 Now by using the Gamma function properties, we have
\begin{eqnarray}\label{57}
\left\{\begin{array}{rr}
\delta_1\varepsilon^k(0)=-\frac{\hbar}{2\sqrt{\pi}}
\lim_{\alpha\rightarrow 0}\mu^{2\alpha}\left[\frac{1}{\sqrt{\pi}\alpha}-\frac{\gamma-\Psi(\frac{1}{2})}{\sqrt{\pi}}+{\cal O} (\alpha)\right],
\\\\\
\delta_1\varepsilon^k(\alpha)=-\frac{\hbar}{\sqrt{\pi}}
\lim_{\alpha\rightarrow 0}\mu^{2\alpha}\left[\frac{1}{2\sqrt{\pi}\alpha}+\frac{\gamma+\Psi(-\frac{1}{2})}{2\sqrt{\pi}}+\cal
O (\alpha)\right],\hspace{.3cm}
\end{array}
\right.
\end{eqnarray}
where $\Psi(z)=\frac{\Gamma^\prime(z)}{\Gamma(z)}$ is digamma
function and $\gamma$ is the Euler-Mascheroni constant . Sum of contributions of two poles leaves a
finite remainder and we end with the finite answer
\begin{eqnarray}\label{58}
\delta_1\varepsilon^k+\delta_2\varepsilon^k=-\frac{m\hbar}{\pi}
\ \ ,\ \ \varepsilon^k=E[\phi_k]-\frac{m\hbar}{\pi}+{\cal O }(
\hbar^{2}\gamma ).
\end{eqnarray}
$$E[\phi_k]= \frac{8m}{\gamma}$$
\begin{eqnarray}\label{59}
\varepsilon^k=\frac{8m}{\gamma}-\frac{m\hbar}{\pi}+{\cal O }(
\hbar^{2}\gamma ).
\end{eqnarray}
The one-loop correction to  SG kink obtained by means
of  generalized zeta function procedure exactly agrees with
accepted result, see [10], [11], [12], [13] and henceforth, with
outcome of the mode number regularization method, [14].
In the case of $\phi^4$-model we left with $l=2$ and then using (\ref{26})
we have
\begin{eqnarray}\label{60}
\int_{-\infty}^\infty \left[G_{\nabla_2}(x,x,\tau)-G_{\nabla_2(0)}(x,x,\tau)\right]dx
=e^{-3\tau}-\frac{3}{\pi}e^{-4\tau}\int_{-\infty}^\infty \frac{(k^2+2)e^{-k^2\tau}}{(k^2+1)(k^2+4)},
\end{eqnarray}
and using (\ref{26}) we obtain
\begin{eqnarray}\label{61}
\begin{array}{cc}
\xi_{\nabla_2}(s)-\xi_{\nabla_2(0)}(s)=3^{-s}-\frac{3}{\pi}\int_{-\infty}^\infty
\frac{dk}{(k^2+4)^{s+1}}-\frac{3}{\pi}\int_{-\infty}^\infty \frac{dk}{(k^2+1)(k^2+4)^{s+1}}=\\
\\
3^{-s}-\frac{3}{\sqrt{\pi}}2^{-(2s+1)}\frac{\Gamma(s+\frac{1}{2})}{\Gamma(s+1)}-\frac{3}{\sqrt{\pi}}2^{-(2s+3)}\frac{\Gamma(s+\frac{3}{2})}{\Gamma(s+2)}
{_2F_1}[s+\frac{3}{2},1,s+2,\frac{3}{4}],
\end{array}
\end{eqnarray}
where we have used the well-known Feynman integral
\begin{eqnarray}\label{62}
\begin{array}{cc}
\frac{1}{D_1^{a_1}D_2^{a_2}...D_n^{a_n}}=\\
\\
\frac{\Gamma(a_1+a_2+...+a_n)}{\Gamma(a_1)\Gamma(a_2)...\Gamma(a_n)}\int
dt_1dt_2...dt_n\frac{\delta(1-t_1-t_2-...-t_n)t_1^{a_1-1}...t_n^{a_n-1}}{(t_1D_1+...+t_nD_n)^{a_1+...+a_n}}.
\end{array}
\end{eqnarray}
Consequently we have
\begin{eqnarray}\label{63}
\begin{array}{cc}
\delta_1\varepsilon^k(s)=\frac{\hbar}{2}\mu^{2s+1}\left[\xi_{P\Delta_2}(s)-\xi_{\Delta_0}(s)\right]=\\
\\
\frac{\hbar}{2}
\left(3^{-s}-\frac{3}{\sqrt{\pi}}2^{-(2s+1)}\frac{\Gamma(s+\frac{1}{2})}{\Gamma(s+1)}-
\frac{3}{\sqrt{\pi}}2^{-(2s+3)}\frac{\Gamma(s+\frac{3}{2})}{\Gamma(s+2)}{_2F_1}[s+\frac{3}{2},1,s+2,\frac{3}{4}]\right).
\end{array}
\end{eqnarray}
Also we obtain
\begin{eqnarray}\label{64}
\begin{array}{cc}
\delta_2\varepsilon^k(s)= \lim_{L\rightarrow \infty}\frac{\hbar}{2L}\mu^{2s+1}\frac{\Gamma(s+1)}{\Gamma(s)}
\xi_{\Delta_2(0)}(s+1)\int_{-\frac{L}{2}}^\frac{L}{2}dx(-6\cosh^{-2}(x))\\
\\
= -\frac{3\hbar}{\sqrt{\pi}2^{2s+1}}\mu^{2s+1}\frac{\Gamma(s+\frac{1}{2})}{\Gamma(s)}.
\end{array}
\end{eqnarray}
Finally we have
\begin{eqnarray}\label{65}
\begin{array}{ccc}
\lim_{s\rightarrow -\frac{1}{2}}(\delta_1\varepsilon^k(s)+\delta_2\varepsilon^k(s))=\frac{\sqrt{3}}{2}\hbar-\frac{3\hbar}{8\sqrt{\pi}}
\frac{\Gamma(1)}{\Gamma(\frac{3}{2})}{_2F_1}[1,1,\frac{3}{2},\frac{3}{4}]\\
\\
-\lim_{\alpha\rightarrow 0} \left(\frac{3\hbar}{\sqrt{\pi}}\frac{\Gamma(\alpha)}{\Gamma(\alpha-\frac{1}{2})}+\frac{3\hbar}{2\sqrt{\pi}}\frac{\Gamma(\alpha)}{\Gamma(\alpha+\frac{1}{2})}\right)\\
\\
=\frac{\hbar}{2\sqrt{3}}-\frac{3\hbar}{\pi}.
\end{array}
\end{eqnarray}
Now using $E[\phi_k]=4m^3/3\lambda$, we find
\begin{eqnarray}\label{66}
\varepsilon^k=\frac{4m^3}{3\lambda}+m\hbar(\frac{1}{2\sqrt{3}}-\frac{3}{\pi}),
\end{eqnarray}
the same answer offered by mode-number regularization method in \cite{R12}.

\section{Conclusion}In this article we used the shape invariance property
of fluctuation operator of SG and $\phi^4$-models to obtain one-loop quantum
correction to the kink mass. This method can be extend to those quantum
fields that their fluctuation operators have shape invariance property. An
interesting extension  worth studying  is to use this method for quantum fields in $(1+2)$-dimension.

{\bf Acknowledgments } The authors would like to thank  H. R. Sepangi for reading the manuscript.

\end{document}